\documentclass[twocolumn,secnumarabic,amssymb, nobibnotes, aps, prd]{revtex4}
\usepackage{graphicx}

\begin{document}
\title{The dc voltage proportional to the persistent current observed on system of asymmetric mesoscopic loops}
\author{V.L. Gurtovoi, A.I. Il'in, A.V.  Nikulov, and V.A. Tulin}
\affiliation{Institute of Microelectronics Technology and High Purity Materials, Russian Academy of Sciences, 142432 Chernogolovka, Moscow District, RUSSIA.} 
\begin{abstract} The observations of the dc voltage proportional to the persistent current on system of asymmetric superconductor loops at a non-zero resistance raise a question on a nature of this quantum phenomenon and its possibility in semiconductor and normal metal mesoscopic loops.
 \end{abstract}

\maketitle

\narrowtext

\section*{Introduction}

It is well known that a potential difference $V = (R_{ls} - R_{l}l_{s}/l)I = R_{an}I$ is observed on a segment $l_{s}$ (with a resistance $R_{ls}$) of an asymmetric conventional metal loop $l$ (with a resistance $R_{l}$) when a circular current $I = \oint_{l} dl E/R_{l}$ is induced by the Faraday's voltage $\oint_{l} dl E = -d\Phi/dt$ in this loop. On the other hand the magnetization measurements give evidence a circular direct current observed in semiconductor [1] normal metal [2] and normal state of superconductor [3] nano-structures in a constant magnetic field, i.e. without the Faraday's voltage $d\Phi/dt = 0$. The observed periodical change of the magnetization with magnetic field at the period corresponding to the flux quantum for single electron $\Phi_{0} = 2\pi \hbar /e$ or pair $\Phi_{0} = \pi \hbar /e$ gives unambiguous evidence that this equilibrium quantum phenomenon, as well as flux quantization in superconductor [4], is a consequence of the persistent current $I_{p}(\Phi/\Phi_{0})$ existing because of the quantization of the velocity circulation 
$$\oint_{l} dl v = \frac{2\pi \hbar }{m}(n  + \frac{\Phi}{\Phi_{0}})  \eqno{(1)}$$ 
But in contrast to the flux quantization observed as far back as 1961 [5] the experimental results [1-3] give evidence of the persistent current along the loop with non-zero resistance $R_{l} > 0$. 

The persistent current at $R_{l} > 0$ was predicted as far bag as 1970 both in normal state $T > T_{c}$ of superconductor [6] and in non-superconductor mesoscopic structures [7]. It was written in [7] and the later theoretical works [8,9] have corroborated that the persistent current can be observed at electron scattering (at a finite mean free path $L_{f.p.}$), i.e. at non-zero dissipation. Thus, the persistent current can be observed at non-zero dissipation like conventional circular current. Nevertheless most experts are fully confident that a potential difference $V_{p}(\Phi/\Phi_{0}) = R_{an}I_{p}(\Phi/\Phi_{0})$ can not be observed on a segment $l_{s}$ when the persistent current $I_{p}(\Phi/\Phi_{0})$ is observed along the asymmetric mesoscopic loop with non-homogeneous dissipation $ R_{an} = R_{ls} - R_{l}l_{s}/l \neq 0$ along its circumference $l$. The observation [10] of the quantum oscillation of the dc voltage $V_{p}(\Phi/\Phi_{0})$ on a system of aluminum loops in the temperature region corresponding to the superconducting resistive transition, i.e. at $R_{l} > 0$, call this confidence in question. 
\begin{figure}[b]
\includegraphics{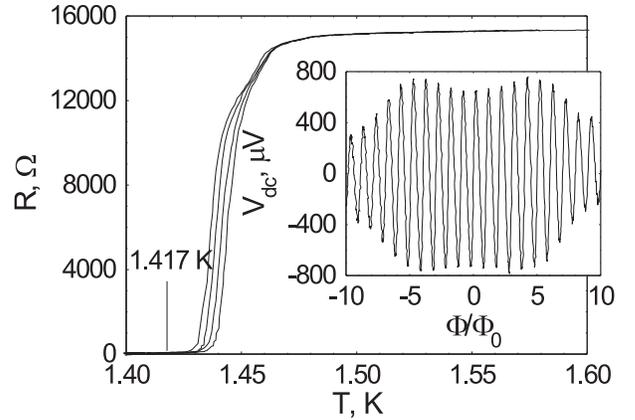}
\caption{\label{fig:epsart} The superconducting resistive transition of the nano-structure containing 1050 asymmetric aluminum loops with diameter $2r = 4 \mu m$ written at the measuring current with different values $I_{ext} = 100 \ nA; 200 \ nA; 300 \ nA; 400 \ nA$. The inset shows the quantum oscillation of the dc voltage $V_{dc}(\Phi/\Phi_{0})$ induced by the external as current with the frequency $f = 1 \ kHz$ and the amplitude $I_{0} = 2 \ \mu A$ at the temperature $T = 1.417 \ K$ corresponding to superconducting state of this nano-structure.}
\end{figure}

\section {The persistent current in superconductor and in non-superconductor loops}
The persistent current observed in normal state of superconductor and non-superconductor (semiconductor and normal metal) has seminar nature and the theorists demonstrate this likeness. I.O. Kulik made the theory of the persistent current in non-superconductor nano-structure [7] just after the work [6] on this phenomenon in normal state of superconductor and in twenty years F. von Oppen and E. K. Riedel have calculated the flux-periodic persistent current in mesoscopic superconducting rings close to $T_{c}$ [11] after the calculation of the disorder-averaged persistent current for a non-superconductor mesoscopic ring [9]. The persistent current can be observed in a loop when the wave function of electron or superconducting condensate is closed now and again in this loop. Therefore the persistent current can have an appreciable value only if the mean free path $L_{f.p.}$ is not smaller than the loop length $l$ [8,9]. 

In the superconducting state the mean free path of pairs is infinite $L_{f.p.} = \infty $ and the persistent current has a value  $I_{p} =  2eN_{s}v_{s}/l$ much large then in a non-superconductor loop $|I_{p}| < ev_{F}/l $ [8,9]. Although the Fermi velocity exceeds the pair velocity $max|v_{s}| = \pi \hbar/ml$ determined by the relation (1) the pair number $N_{s}$ in any real loop is so great at $T < T_{c}$ that $2eN_{s}\pi \hbar /ml^{2} \gg ev_{F}/l $. Because of the large $I_{p}$ value the quantum oscillation of the dc voltage $V_{dc}(\Phi/\Phi_{0})$ with high amplitude can be observed at $T < T_{c}$, Fig.1. But because of zero resistance $R_{an} = 0$ an external ac current with the amplitude $I_{0}$ exceeding the superconducting critical current $I_{c} = I_{c}(0)(1-T/T_{c})^{3/2}$ should be applied at $T < T_{c}$ [12]. 

\begin{figure}
\includegraphics{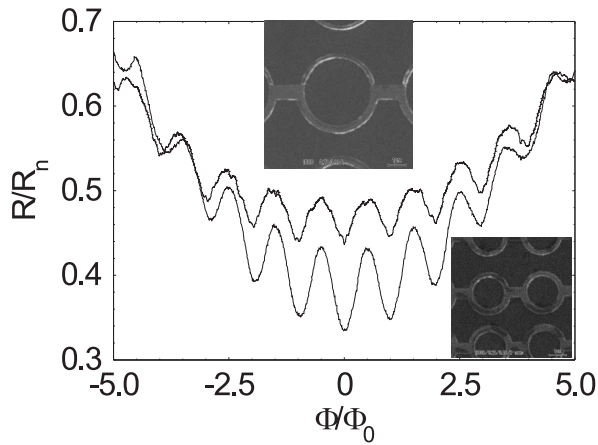}
\caption{\label{fig:epsart} The Little-Parks oscillations of the resistance $R$ reduced to the one in the normal state $R_{n}$ measured on two nano-structures containing aluminum loops with diameter $2r = 4 \mu m$ (the upper curve) and $2r = 2 \mu m$ (the lower curve) demonstrate the increase of the amplitude of the superconducting transition shift $\Delta T_{c}(\Phi/\Phi_{0})$ in magnetic field with the loop decrease. }
\end{figure}

\section{Shift because of the persistent current and width of superconducting resistive transition}
Such switching between quantum states with different connectivity of the wave function can induce a potential difference $V_{p}(\Phi/\Phi_{0}) \propto I_{p}(\Phi/\Phi_{0})$ on segment of an asymmetric loop [14,15]. It is expected that its value in the normal state $T > T_{c}$ may be larger than in non-superconductor loop since the $I_{p}$ value in the first case [3] is larger than in the second one [1,2]. The persistent current $I_{p} \propto v_{s} \propto 1/l$ increases with the loop length $l = 2\pi r$ decrease. But at a too small loop $r < \xi(0)(\delta T_{c}/\Delta T_{c,sh})^{1/2}$ the switching between states with different connectivity of the wave function becomes impossible [14] because of the critical temperature shift $\Delta T_{c} = \Delta T_{c,sh}(n - \Phi/\Phi_{0})^{2} =  -(\xi(0) /r)^{2}(n - \Phi/\Phi_{0})^{2}$ [16]. Here $\xi(0)$ is the superconductor coherence length at $T = 0$ and $\delta T_{c}$ is the width of the superconducting transition. Our measurements have corroborated the $\Delta T_{c}(\Phi/\Phi_{0}) \propto \Delta R(\Phi/\Phi_{0})/R_{n}$ amplitude increase with the $2r$ loop decrease, Fig.2. We have found that $\Delta T_{c,sh} =  \delta T_{c}$ at the diameter of our aluminum loop $2r = 2 \ \mu m$. We intend to present results of the $V_{p}(\Phi/\Phi_{0})$ measurements on nano-structures with great number of such loops, Fig.3. It may be these results will answer on the question on a possibility to observe the like phenomenon in semiconductor and normal metal loops.  

\begin{figure}[]
\includegraphics{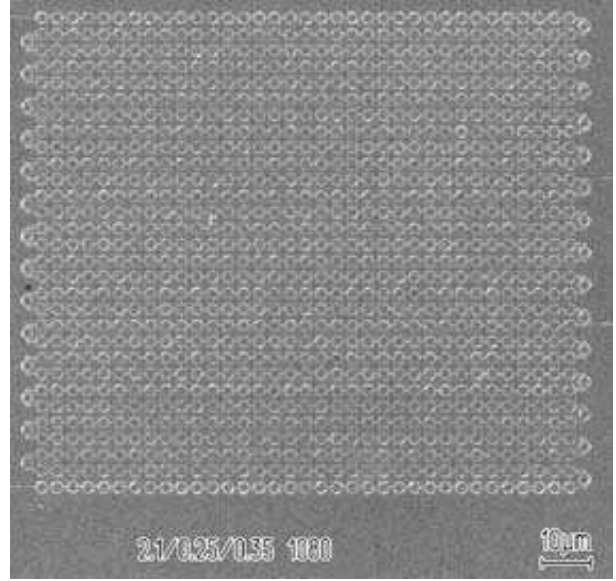}
\caption{\label{fig:epsart} An electron micrograph of the nano-structure containing 1080 asymmetric aluminum loops with diameter $2r = 2 \mu m$.}
\end{figure} 

\section*{Acknowledgement}
This work has been supported by grant of the Program "Quantum Nanostructures" of the Presidium of RAS, grant 08-02-99042-r-ofi of the  Russian Foundation of Basic Research and grant "Quantum bit on base of micro- and nano-structures with metal conductivity" of the Program "Technology Basis of New Computing Methods" of ITCS department of RAS.

\end{document}